\documentclass[12pt,pre,aps,superscriptaddress]{revtex4}
\usepackage{graphicx}
\newcommand{\beq}[2]{\begin{equation}#1\label{#2}\end{equation}}
\newcommand{\ceq}[1]{(\ref{#1})}

\newfont{\mbld}{cmbx10 scaled 800}
\newfont{\cab}{cmsy10 scaled 1200}
\newfont{\scab}{cmsy10 scaled 1000}
\newfont{\bcall}{cmbsy10 scaled 1200}

\begin{document}
\title{A path integral approach to the dynamics of random chains}
\author{Franco Ferrari}
\email{ferrari@univ.szczecin.pl}
\author{Jaros{\l}aw Paturej}\email{jpaturej@univ.szczecin.pl}
\affiliation{Institute of Physics and CASA*, University of Szczecin,
  ul. Wielkopolska 15, 70-451 Szczecin, Poland}
\author{Thomas A. Vilgis} 
\email{vilgis@mpip-mainz.mpg.de}
\affiliation{Max Planck Institute for Polymer Research, 10
  Ackermannweg, 55128 Mainz, Germany}

\begin{abstract}
  In this work the dynamics of a freely jointed random chain with small masses
  attached to the joints is studied from a microscopic point of view.  The
  chain is treated using a stringy approach, in which a statistical sum is
  performed over all two dimensional trajectories spanned by the chain during
  its fluctuations.  In the limit in which the chain becomes a continuous
  curve, the probability function for such a system coincides with the
  partition function of a generalized nonlinear sigma model. The cases of open
  or closed chains in two and three dimensions are discussed. In three
  dimensions it is possible also to introduce some rigidity at the joints,
  allowing the segments of the chain to take only particular angles with
  respect to a given direction.
\end{abstract}
\maketitle
\section{Introduction}\label{sec:intro}
In this paper the dynamics of a random chain subjected to thermal fluctuations
at fixed temperature $T$ is discussed.  This problem is usually treated
phenomenologically, regarding the fluctuations of the chain as a stochastic
process which may be described with the help of Langevin equations or,
alternatively, of Fokker-Planck equations \cite{doiedwards}. This approach
leads to the well known models of Rouse \cite{rouse} and Zimm \cite{zimm}
which allow a satisfactory understanding of the main properties of polymers in
solutions. One major drawback of these coarse grained models is, that they
suffer from the presence of rigid constraints. The Rouse and Rouse--Zimm
equation consider only chains bead spring models of chains, where the local
spring is infinitely extensible. In the framework of continuous models the
Rouse equation for example is nothing but the stochastic equation (Langevin
equation) for the classical Wiener measure, which yields paths which do not
have a well defined tangent vectors \cite{doiedwards}. These problems have
been tackled by various attempts, see e.g. \cite{yamakawa,stockmayer}.
However, the correct use of rigid constraints in (stochastic) dynamics requires
some mathematical effort \cite{arti, zinn}, in contrast to the static cases
where rigid constraints can be implemented by Dirac delta functions in the
partition function.

However, it seems to be too difficult to extend these approaches, therefore we
follow here some different routes. Moreover the use of dynamic constraints for
(stochastic dynamics equation) is always problematic. 

Here the dynamics of random chains is considered from a microscopic point of
view. The chain is represented as a set of freely jointed segments of fixed
lengths. Small masses are attached at the joints.  The goal is to construct
the probability function $\Psi[\mbox{Conf}_f,\mbox{Conf}_i]$ of this system.
Roughly speaking this function measures the probability that a chain starting
from a given initial conformation Conf$_i$ at the time $t_i$ arrives to a
given final conformation Conf$_f$ at time $t_f$.  During its motion from the
initial to the final conformation the chain spans a two dimensional surface.
In this way one obtains a {\it stringy} formulation of the chain dynamics,
which is based on path integrals \cite{kleinertpi}.  The chain ``world-sheet''
is a rectangle whose sides are given by the length of the chain and the time
interval $\Delta t=t_f-t_i$.

The basic ingredient in the construction of the probability function
$\Psi[\mbox{Conf}_f,\mbox{Conf}_i]$ is the energy of a discrete freely jointed
chain with $N$ segments.  While it is easy to add external and internal
interactions acting on the segments of the chain, the complicated form of the
kinetic energy poses serious obstacles to the possibility of computing the
probability function analytically.  The kinetic energy is in fact nonlinear
and has an intrinsically non-Markoffian nature. This is due to the fact that
the velocity of each segment in the chain depends on the velocities of the
other segments.  In this situation, it is difficult to isolate in the energy a
linear term which could give raise to a propagator, so that even perturbative
calculations are not allowed.  Luckily, the expression of the kinetic energy
simplifies considerably after performing the limit in which the chain becomes
a continuous curve. The final form of the probability function which one
obtains in this limit closely resembles the partition function of a nonlinear
sigma model \cite{nlsigma}.  The difference with respect to the standard
nonlinear sigma model is that in the latter case the modulus of the fields is
constrained, while in the present case the constraint involves the modulus of
the derivative of the fields with respect to the arc-length of the chain.
This constraint is related to our assumption that the lengths of the segments
is fixed.

Within our formalism it is possible to add constraints to the trajectory of
the chain with the help of Dirac $\delta-$functions as in the case of the
statistical mechanics of random chains \cite{edwa}. Here we have considered
just the simplest example of constraints, namely the requirement that the
trajectory of the chain is closed.  Besides, it is easy to treat
non-homogeneous chains, in which both lengths of the segments and the masses
located at the points are arbitrary.  The probability functions is constructed
for two and three dimensional random chains.  In three dimensions one may
introduce further constraints, which fix for instance the lengths of the
projections of the segments on the $z$ axis. In this way we are able to
discuss also rigid chains, in which the segments are allowed to form only a
given angle with respect to the $z$ axis.

The material presented in this paper is divided as follows.  In Section II the
expression of the kinetic energy of a discrete chain with $N$ segments is
derived in two dimensions using a recursive method. The rules of the passage
to the continuous limit are established. After this limit is performed,
important simplifications occur.  To further simplify the problem, 
one of the ends of the chain is fixed at a given point.
Moreover, it is supposed that the distributions of masses and segment lengths
along the chain are uniform.  It is shown that under these assumptions
the kinetic energy of the
continuous chain may be written in terms of free complex scalar fields
subjected to a constraint. The origin of this constraint is the requirement
that the lengths of the segments are constant.  In Section III the classical
solutions of a free chain are studied. In Section IV a stringy approach to the
dynamics of the chain in two dimensions is established. The probability
function is constructed using path integrals.  The resulting model closely
resembles a nonlinear supersymmetric model.  The problem of fixing various
boundary conditions, including the case of closed chain trajectories, is
discussed. A perturbative approach based on the expansion of the fields
describing the statistical fluctuations over a classical background is
provided. In Section V the results of Sections II--IV are extended to three
dimensional chains. In Section VI chains with constant bending angles are
investigated. Finally, in Section VII our conclusions are presented.

\section{The energy of a free chain}\label{sec2}

\begin{figure}[bpht]
\centering
\includegraphics[width=.5\textwidth]{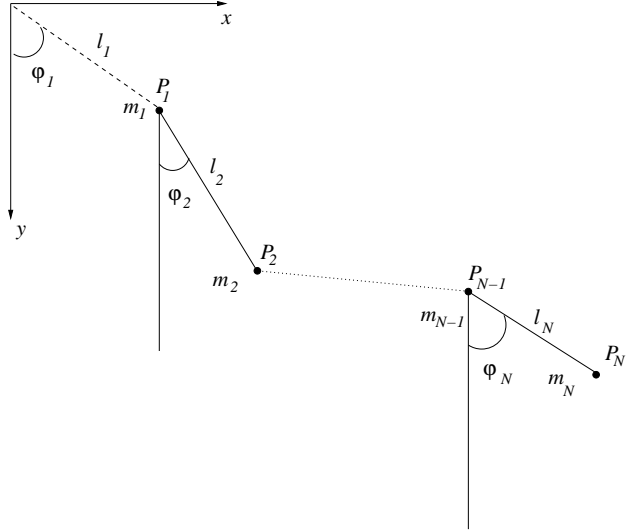}
\caption{\footnotesize
A chain with $N$ segments. Let us note that the end point
  $P_1$ is not bound to stay at a fixed distance with respect to the
  origin of the cartesian reference system.
}\label{system2y}
\end{figure}

Let us consider a chain of $N$ segments of fixed lengths $l_2,\ldots,l_N$ in
the two dimensional plane. Each segment $P_{i+1}P_i$ is completely specified
by the positions of its end points $P_{i+1}$ and $P_i$. In cartesian
coordinates $(x,y)$ these positions are given by the radius vectors:
\beq{\mathbf r_i=(x_i,y_i)\qquad\qquad i=1,\ldots,N}{pointcoor} The segments
are joined together at the points $P_l$, where $2\le l\le N-1$, see
Fig.~\ref{system2y}, while $P_1$ and $P_N$ are the ends of the chain.
Moreover, at each point $P_i$, with $i=1,\ldots,N$, a mass $m_i$ is attached.
In the following we restrict ourselves to the case of a free chain. We will
see below that the addition of interactions is straightforward.

To compute the kinetic energy of the above system, it is convenient to pass to
polar coordinates $l_i,\varphi_i$ as follows: 

\beq{
  x_n=\sum_{i=1}^nl_i\cos\varphi_i\qquad\qquad y_n=\sum_{i=1}^n
  l_i\sin\varphi_i\qquad\qquad(n=1,\ldots,N) }{carpol2} The $\varphi_i$ is the
angle formed by segment $i$ with the $y-$axis, see Fig.~\ref{system2y}.  All
the radial coordinates $l_i$ are constants for $i=2,\ldots,N$. The only
exception is the length $l_1(t)$ which denotes the distance of the point
$x_1,y_1$ from the origin. Since this distance is not fixed, $l_1=l_1(t)$ is
allowed to vary with the time $t$.  From Eq.~\ceq{carpol2} the velocity
components of the $n-$th segment may be written as follows:
\begin{eqnarray}
\dot
x_n&=&-\sum_{i=1}^{n-1}l_i\dot\varphi_i\sin\varphi_i-l_n
\dot\varphi_n\sin\varphi_n   
+\dot l_1\cos\varphi_1
\qquad\qquad(n=2,\ldots,N)\label{velcom1} \\
\dot
y_n&=&\sum_{i=1}^{n-1}l_i\dot\varphi_i\cos\varphi_i+l_n
\dot\varphi_n\cos\varphi_n   
+\dot l_1\sin\varphi_1
\qquad\qquad(n=2,\ldots,N) \label{velcom2}\\
\dot x_1&=&-l_1\dot\varphi_1\sin\varphi_1+\dot
l_1\cos\varphi_1\label{velcom3}\\
\dot y_1&=&l_1\dot\varphi_1\cos\varphi_1+\dot
l_1\sin\varphi_1\label{velcom4}
\end{eqnarray}
Separating the contribution coming from the first $n-1$ variables, the kinetic
energy $K_n$ of the $n-$th segment can be expressed in terms of the kinetic
energy $K_{n-1}$ of the $(n-1)-$th segment: 
\beq{
  K_n=\frac{m_n}{m_{n-1}}K_{n-1}+\frac{m_n}{2}l_n^2\dot\varphi_n^2+
  m_n\sum_{i=1}^{n-1}l_nl_i\dot\varphi_n\dot\varphi_i\cos(\varphi_i-\varphi_n)
  +m_nl_n\dot\varphi_n\dot l_1\sin(\varphi_1-\varphi_n) }{kinenerecrel} It is
possible to solve the above recursion relation to find an expression of $K_n$.
If we do that, at the end the total kinetic energy of the discrete chain:
\beq{K_{disc}^{2d}=\sum_{n=1}^NK_n}{totenechadef} becomes:
\begin{eqnarray}
K_{disc}^{2d}&=&\frac M2\left(
l_1^2\dot\varphi_1^2+\dot l_1^2
\right )
+l_1\dot\varphi_1\sum_{n=1}^N
\sum_{k=1}^{n-1}m_n
l_{n-k+1}
\dot\varphi_{n-k+1}\cos(\varphi_{n-k+1}-\varphi_1)
\nonumber\\  
&+&\dot l_1\sum_{n=1}^N
\sum_{k=1}^{n-1}m_nl_{n-k+1}\dot\varphi_{n-k+1}
\sin(\varphi_1-\varphi_{n-k+1})\nonumber\\
&+&
\sum_{n=1}^N
\sum_{k=1}^{n-1}l_{n-k+1}^2\frac{m_n}2\dot\varphi_{n-k+1}^2
+\sum_{n=1}^N\sum_{k=1}^{n-1}\sum_{i=2}^{n-k}m_nl_{n-k+1}l_i
\dot\varphi_{n-k+1} \dot\varphi_i\cos(\varphi_{n-k+1}-\varphi_i)\label{tdiscr}
\end{eqnarray}
where $M=\sum_{n=1}^Nm_n$ is the total mass of the chain \footnote{Our
result agrees with that of Ref.~\cite{tomapier}, where a similar
calculation has been recently reported. Our expression of the kinetic
energy of the chain is slightly more general, since both ends of the
chain are free to move.}.

We wish now to perform the limit in which the chain of $N$ segments
becomes a continuous system \footnote{An exhaustive discussion about
  the passage from discrete to continuous random chain systems may be
  found in \cite{kleinertgf}.}. To this purpose, it is convenient to
consider the indices $i,k,n,\ldots$ appearing in Eq.~\ceq{tdiscr} as
discrete variables taking values in a one dimensional lattice with $N$
points. Quantities $f_i$ carrying the index $i$ may be interpreted as
functions of $i$. Their variations $\Delta f_i$ are given by: $\Delta
f_i=f_{i+1}-f_i$. Clearly,  $\Delta_ii=1$, i. e. the
spacing between two neighboring points in the lattice is 1. In order
to proceed, we rescale the distances in the lattice in such a way that
the spacing in the new lattice will be $a$. To this purpose, we
perform the transformations $i\longrightarrow s_i$,
$f_i\longrightarrow f(s_i)$ where the new variable $s_i$ has variation
$\Delta s_i=s_{i+1}-s_i=a$. The next step is to compute the kinetic
energy of Eq.~\ceq{tdiscr} in the limit $N\longrightarrow\infty$,
$a\longrightarrow 0 $, while the product $Na$ remains finite, let's
say $Na=L$, where $L$ denotes the total length of the chain. Clearly,
in this limit the right hand side of Eq.~\ceq{tdiscr} will diverge
unless we suppose that the masses $m_i$ and the lengths $l_i$ of the
segments are going to zero in a suitable way.
Reasonable assumptions are:
\beq{l_i\longrightarrow l(s_i)=a\rho_l(s_i)\qquad\qquad
m_i\longrightarrow m(s_i)=a\rho_m(s_i)}{ansaone}
where $\rho_l(s_i)$ and $\rho_m(s_i)$ are respectively the
distribution of length and of mass within the chain. Here we allow for
segments of different lengths $l(s_i)$ and for points $P_i$ of
different masses $m(s_i)$.
To be consistent with our settings, the distributions $\rho_l(s_i)$
and $\rho_m(s_i)$ must be normalized as follows:
\beq{
\sum_{i=1}^N\rho_l(s_i)\Delta s_i=L\qquad\qquad
\sum_{i=1}^N\rho_m(s_i)\Delta s_i=M
}{nordist}
At this point we are ready to pass to the continuous limit. Functions
of discrete variables will be substituted with functions of continuous
variables, while sums will be replaced with integrals according to the
following rules:
\beq{f(s_i)\longrightarrow f(s)\qquad\qquad \sum_{i=1}^N\Delta
  s_i\longrightarrow \int_0^Lds}{disccontrul}
After a few calculations one finds:
\beq{K_{disc}^{2d}\longrightarrow K^{2d}}{kdiscik}
where
\begin{eqnarray}
K^{2d}&=&K^{2d}(t)=\frac M2(l_1^2(t)\dot\varphi_1^2(t)+\dot
l_1^2(t))\nonumber\\ 
&\!\!\!\!\!\!\!\!+&\!\!\!\!\!\!\dot\varphi_1(t)l_1(t)\int_0^Lds\rho_m(s)
\int_0^s
du\rho_l(s-u)\dot\varphi(t,s-u)\cos(\varphi(t,s-u)-\varphi_1(t))\nonumber
\\
&\!\!\!\!\!\!\!\!+&\!\!\!\!\!\!\dot
l_1(t)\int_0^Lds\rho_m(s)\int_0^Ldu\rho_l(s-u)\dot\varphi(t,s-u)\sin(
\varphi_1(t) -\varphi(t,s-u))\nonumber\\
&\!\!\!\!\!\!\!\!+&\!\!\!\!\!\!\!\!\!\int_0^Lds\rho_m(s)\int_0^sdu\rho_l(s-u)
\int_0^{s-u}\!\!\!dv 
\rho_l(v)\dot\varphi(t,s-u)  
\dot \varphi(t,v) \cos(\varphi(t,s-u)-\varphi(t,v))
\label{Tcont}
\end{eqnarray}
and
\beq{
\int_0^Lds\rho_l(s)=L\qquad\qquad\int_0^Lds\rho_m(s)=M
}{addpos}
Eq.~\ceq{Tcont} may be simplified by performing in the integrals in
$du$ the following change of variables:
\beq{
u'=s-u\qquad\qquad du'=-du
}{chavar}
and then using the formula:
\beq{
\int_0^Lds\int_0^sdu'f(u')=\int_0^Lds(L-s)f(s)
}{usefor}
which is valid for any integrable function $f(s)$.
As a result, we obtain:
\begin{eqnarray}
&K^{2d}&=\frac M2(l_1^2(t)\dot\varphi_1^2(t)+\dot l_1^2(t))\nonumber\\
&+&\dot\varphi_1(t)l_1(t)\int_0^Lds(L-s)\rho_m(s)
\rho_l(s)\dot\varphi(t,s)\cos(\varphi(t,s)-\varphi_1(t))\nonumber
\\
&+&\dot
l_1(t)\int_0^Lds(L-s)\rho_m(s)\rho_l(s)\dot\varphi(t,s)\sin(
\varphi_1(t) -\varphi(t,s))\nonumber\\
&+&\!\!\!\int_0^Lds(L-s)\rho_m(s)\rho_l(s)\int_0^sdu
\rho_l(u)\dot\varphi(t,s)  
\dot\varphi(t,u) \cos(\varphi(t,s)-\varphi(t,u))
\label{Tcontsimp}
\end{eqnarray}
For simplicity, we suppose the length and mass distributions in the
chains are uniform. As a consequence, we put:
\beq{
\rho_l(s)=1\qquad\qquad \rho_m(s)=\frac ML
}{unifass}
The first of Eqs.~\ceq{unifass} implies (see Eq.~\ceq{ansaone}):
\beq{l_i=a\qquad\qquad i=2,\ldots,N}{liallequal}
Remembering the definition of the length distribution in
Eq.~\ceq{ansaone}, it is easy to realize that the first of 
Eqs.~\ceq{unifass} implies that all segments of the chain have the
same length. 
As a further simplification, we will study the case in which the point
$P_1$ is fixed, so that 
\beq{
\dot l_1=\dot\varphi_1=0
}{ponefixass}
From the above assumptions and from Eq.~\ceq{Tcontsimp}, we find that
the total energy ${\cal H}_0$ of the ideal chain is given by:
\beq{
{\cal H}_0(\varphi)=\frac ML\int_0^Lds
(L-s)\int_0^sdu\dot\varphi(t,s)\dot\varphi(t,u)
\cos(\varphi(t,s)-\varphi(t,u))
}{hamfrecha}
In view of a future path integral treatment of the dynamics of the
chain, it will be convenient to 
use cartesian coordinates, because of the complications involved with
polar coordinates in the path integral approach.
To this purpose, we
rewrite Eq.~\ceq{hamfrecha} with the help of the well known
trigonometric formula: 
\beq{
\cos(\varphi(t,s)-\varphi(t,u))=\frac12\left(
e^{i\varphi(t,s)}e^{-i\varphi(t,u)}+e^{-i\varphi(t,s)}e^{i\varphi(t,s)}
\right)
}{noteone}
Introducing new complex variables
\begin{eqnarray}
\Phi(t,s)=\int_0^sdue^{i\varphi(t,u)}+l_1e^{i\varphi_1}\label{zst}\\
\bar \Phi(t,s)=\int_0^sdue^{-i\varphi(t,u)}+l_1 e^{-i\varphi_1}
\label{zbst}\\
\end{eqnarray}
one finds after simple calculations:
\beq{
{\cal H}_0(\varphi)=\frac M{2L}\int_0^Lds\frac{\partial\bar
  \Phi(t,s)}{dt}
\frac{\partial \Phi(t,s)}{dt}
}{szerocomp}
Now the expression of the energy has become simpler, but the
new fields
$\Phi(s,t)$ and $\bar \Phi(s,t)$ are not independent and have a
complicated
dependence on the true degree of freedom
$\varphi(t,s)$ as shown by Eqs.~\ceq{zst} and \ceq{zbst}.
For this reason, it is preferable to assume that $\Phi(t,s)$ and
$\bar\Phi(t,s)$ are independent complex fields subjected to the
constraint 
\footnote{Let us note that in Eq.~\ceq{constr} and from now on we will
  use the following 
notations to denote partial derivatives in $t$ and $s$:
$$
\frac{\partial f(t,s)}{\partial t}=\partial_t f(t,s)=\dot f(t,s)
\qquad\qquad
\frac{\partial f(t,s)}{\partial s}=\partial_s f(t,s)=f'(t,s)
$$
}:
\beq{
\partial_s\bar \Phi(t,s)\partial_s \Phi(t,s)=1
}{constr}
It is easy to check that Eqs.~\ceq{zst} and \ceq{zbst} provide
exactly the solution of the constraint \ceq{constr}.
At this point the energy ${\cal H}_0(\varphi)$ may be rewritten 
in the simple form:
\beq{
{\cal H}_0(\Phi,\bar \Phi)=\frac M{2L}\int_0^Lds\frac{\partial\bar
  \Phi(t,s)}{dt}
\frac{\partial \Phi(t,s)}{dt}
}{szerocompindep}
where $\Phi,\bar \Phi$ are treated as independent complex degrees of freedom
  subjected to the condition \ceq{constr}. This constraint can be
  imposed for instance by means of a Lagrange multiplier.

Instead of the complex coordinates $\Phi,\bar\Phi$ one may also
exploit real coordinates $x,y$:
\begin{eqnarray}
x(t,s)=\int_0^sdu\cos\varphi(t,u)+l_1\cos\varphi_1
=\frac12(\Phi(t,s)+\bar
\Phi(t,s))
\nonumber\\
y(t,s)=\int_0^sdu\sin\varphi(t,u)+l_1\sin\varphi_1=\frac1{2i}(\Phi(t,s)-\bar
\Phi(t,s))
\label{realvar} 
\end{eqnarray}
It is easy to see
that $x(t,s)$ and $y(t,s)$
correspond in the continuous case to the discrete coordinates
$x_n$ and $y_n$ of the point $P_n$ given by Eq.~\ceq{carpol2}.
Written in terms of the real variables
$x(t,s)$ and $y(t,s)$ the functional ${\cal H}_0(\Phi,\bar \Phi)$
and the constraint \ceq{constr}
become respectively:
\beq{{\cal H}_0(x,y)=
\frac{M}{2L} \int_0^Lds(\dot x^2(t,s)+\dot y^2(t,s))
}{hzeroreal}
and
\beq{(\partial_sx(t,s))^2+(\partial_sy(t,s))^2=1
}{constrxy}
In the discrete case this constraint corresponds to
the condition:
\beq{
(x_i-x_{i-1})^2+(y_i-y_{i-1})^2=a^2\qquad\qquad i=2,\ldots,N
}{constrdisc}

In the future it will
be convenient to exploit the vector notation:
\beq{
\mathbf R(t,s)=(x(t,s),y(t,s))
}{vecnot}
and
\beq{{\cal H}_0(x,y)={\cal H}_0(\mathbf R)
=\frac{M}{2L} \int_0^Lds\dot \mathbf R^2(t,s)
}{dffdf}
If $s=0$, it is clear from Eq.~\ceq{realvar}
that the point $\mathbf R(t,0)$ is fixed for every time $t$ at the
location:
\beq{\mathbf R(t,0)=(l_1\cos\varphi_1,l_1\sin\varphi_1)}{begbb}
This is in agreement with the
assumption of Eq.~\ceq{ponefixass}, where it is supposed that
$l_1$ and $\varphi_1$ are constant in time.
\section{The classical equations of motion}\label{sec:class}
We wish to give a ``stringy'' interpretation of the chain dynamics, in
which the chain moves during a time interval $[t_i,t_f]$ from an
initial conformation $\Phi_i(s)$ to a final conformation
$\Phi_f(s)$. During this motion, the chain spans a two dimensional
portion of the plane $(x,y)$, whose points are described by the
``complex coordinates'' $\Phi(t,s),\bar\Phi(t,s)$.
The chain ``world-sheet'' is delimited by the range of the variables
$t$  and $s$: $t_i\le t\le
t_f$ and $0\le s\le L$. The situation is depicted in Fig.~\ref{dynamic}.
Following this strategy, we define the chain action:
\beq{{\cal A}_0=
\frac M{2L}\int_{t_i}^{t_f}dt\int_0^Lds\frac{\partial\bar
  \Phi(t,s)}{dt}
\frac{\partial \Phi(t,s)}{dt}
}{actionzerodef}
In the rest of this Section we study the classical equations of motion
corresponding to the action
${\cal A}_0$:
\beq{
\frac{\partial^2 \Phi(t,s)}{\partial t^2}=
\frac{\partial^2 \bar \Phi(t,s)}{\partial t^2}=0
}{clone}
and the constraint \ceq{constr}:
\beq{
\left|\frac{\partial \Phi(t,s)}{\partial s}\right|^2=1
}{cltwo}
The solutions of Eqs.~\ceq{clone} are of the form:
\beq{\Phi_{cl}(t,s)=a(s)+tb(s)\qquad\qquad \bar \Phi_{cl}(t,s)=\bar
  a(s)+t\bar b(s) 
}{solcla}
$a$ and $b$ are complex functions of $s$ and $\bar a,\bar b$ are their
complex conjugates.
If we put
\beq{
a(s)=\frac{t_f\Phi_i(s)-t_i\Phi_f(s)}{t_f-t_i}
\qquad\qquad b(s)=\frac{1} {t_f-t_i }(\Phi_f(s)-\Phi_i(s))
}{clatra}
then Eq.~\ceq{solcla} represents the evolution of
a chain which during the time $t_f-t_i $ passes
from an initial conformation $\Phi_i(s)$ to a final conformation
$\Phi_f(s)$ for $0\le s\le L$.
The constraint \ceq{constr} requires additionally that $b$ and $\bar
b$ are constants independent of $s$ and that:
\beq{
\partial_sa(s)=e^{i\tilde\varphi(s)}\qquad\qquad
\partial_s\bar a(s)=e^{-i\tilde\varphi(s)}
}{afterconstraint}
where $\tilde\varphi(s)$ describes the angles of  a given static
conformation of the chain.
Going to the real coordinates $x(t,s)$ and $y(t,s)$ of
Eq.~\ceq{realvar}, this implies that:
\begin{eqnarray}
x_{cl}(t,s)\equiv x_{cl}(s)=
\int_0^sdu\cos\tilde\varphi(u)+l_1\cos\varphi_1\label{xclone}\\
y_{cl}(t,s)\equiv y_{cl}(s)=
\int_0^sdu\sin\tilde\varphi(u)+l_1\sin\varphi_1\label{xcltwo}
\end{eqnarray}
We note that in the above equation we have put $b=\bar b=0$, so that
there is no dependence on the time.
This is required by condition \ceq{begbb}, which
demands that the beginning of
the chain is fixed at the point $(l_1\cos\varphi_1,l_1\sin\varphi_1)$.
In other words, in the absence of interactions, the conformation of
the chain does not change in time.
This is not surprising.
In fact, we note that, in the passage from the
discrete kinetic energy $K_{disc}^{2d}$ to its continuous counterpart,
the term $\sum_{n=1}^N
\sum_{k=1}^{n-1}l_{n-k+1}^2\frac{m_n}2\dot\varphi_{n-k+1}^2$ which was
present in $K_{disc}^{2d}$ disappeared after performing
the limit
$a\longrightarrow 0$. This fact, together with the conditions
\ceq{ponefixass}, which fix one of the ends of the chain, make the
classical dynamics of the ideal chain trivial.

As anticipated in the previous Section, it is now easy to add the
interactions. For example, let us suppose that the segments of the
chain are immersed in an external potential $V_{ext}(\mathbf r)$ and
that there are also internal interactions associated to a two-body
potential $V_{int}(\mathbf r_1,\mathbf r_2)$.
In this case, Eq.~\ceq{szerocomp} generalizes to:
\beq{
{\cal A}={\cal A}_0+{\cal A}_{ext}+{\cal A}_{int}
}{sgen}
where ${\cal A}_0$ has been defined in Eq.~\ceq{actionzerodef}, while
\beq{
{\cal A}_{ext}=\int_{t_i}^{t_f} dt\int_0^Lds V_{ext}(\mathbf R(t,s))
}{eneext}
and
\beq{
{\cal A}_{int}=\int_{t_i}^{t_f} dt_1\int_0^Lds_1
\int_{t_i}^{t_f} dt_2\int_0^Ld s_2V_{int}(
\mathbf R(t_1,s_1),\mathbf R(t_2,s_2)
)
}{eneint}
Alternatively, if one wishes to express ${\cal A}_{ext}$ and ${\cal
  A}_{int}$ as functionals of 
$\Phi(t,s)$ and $\bar \Phi(t,s)$ it is possible to exploit Eqs.~\ceq{realvar}.
\section{Dynamics of a chain immersed in a thermal bath}
At this point we are ready to study the dynamics of a chain
fluctuating in a solution at fixed temperature $T$. Once the
action of the system is known, it is possible to introduce the
dynamics using a path integral approach. To this purpose, we consider
the following probability distribution:
\beq{
\Psi^{2d}
=\int_{
\Phi(t_f,s)=\Phi_f(s)\atop
\Phi(t_i,s)=\Phi_i(s)
}{\cal D}\Phi(t,s){\cal D}\bar{
\Phi}(t,s) \exp\left({-\displaystyle
\frac{\cal A}{k_BT}}\right)\delta(|\partial_s \Phi(t,s)|^2-1)
}
{psifixlen}
where ${\cal A}$ is the action of
Eqs.~(\ref{sgen})--(\ref{eneint}).
The
boundary conditions for the complex conjugated field $\bar \Phi$
do not appear in Eq.~\ceq{psifixlen}, because they
are fixed by the boundary conditions of the field $\Phi$
due the constraint \ceq{constr}, which is imposed in
Eq.~\ceq{psifixlen} by means of the Dirac $\delta-$function.
Let us also note that due to that constraint
the action ${\cal A}$ may be written in a form which
resembles more that of a two dimensional field theory:
\beq{
{\cal A}=\frac{M}{2L}\int_{t_i}^{t_f} dt\int_0^Lds|\mathbf\nabla \Phi(t,s)|^2+
{\cal A}_{ext}+{\cal A}_{int}
}{actftform}
with $\mathbf \nabla=(\partial_t,\partial_s)$. 
The actions ${\cal A}_{ext}$ and ${\cal A}_{int}$ are useful in order
to describe the interactions of the chain with itself and with the
surrounding environment.
\begin{figure}
\centering
\includegraphics[width=8cm]{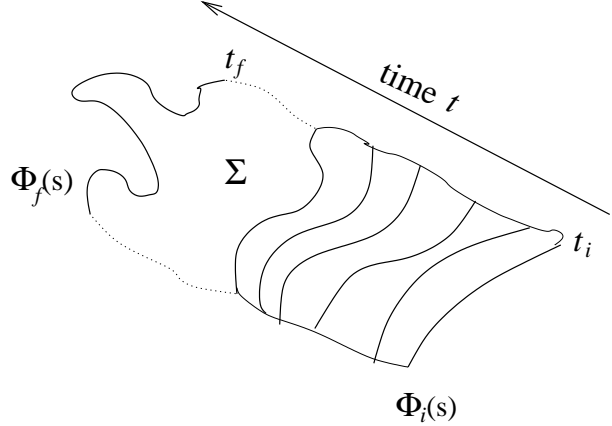}
\caption{In its motion from the initial conformation $\Phi_i(s)$ to
  the final conformation $\Phi_f(s)$ the chain spans a
  two dimensional surface in the space.} \label{dynamic}
\end{figure}
The distribution
$\Psi^{2d}
$ gives the
probability that a chain of length $L$
starting from an initial conformation $\Phi_i(s)$ at time $t=t_i$
arrives to a
final conformation $\Phi_f(s)$ at the instant $t=t_f$ 
after fluctuating in a thermal bath held at constant
temperature
$T$.
To understand how this probability is computed by means of the path
integral appearing in the right hand side of Eq.~\ceq{psifixlen} it is
helpful to look at
Fig.~\ref{dynamic}
\footnote{For simplicity, in Fig.~\ref{dynamic}  it has been
  described
 a
  chain in which both ends 
are allowed to move. We should however remember
that, from our construction, one end of the chain remains
always fixed according to Eq.~\ceq{begbb}.}. 
During its motion in the 
time interval $[t_i,t_f]$ the
chain spans a surface $\Sigma$. In
Eq.~\ceq{psifixlen} it is performed a sum
 over all possible surfaces of this type. If
the temperature  $T$ is zero, the chain  moves
according to the classical equations of motion. In that case, the
conformations of minimal energy are favored and the energy
 is conserved along the motion at each instant. If however the
temperature is different from zero, the energy does not need to be
conserved in time, because of the thermal fluctuations. The
conformations for which the energy is not minimal are suppressed in
the exponential $\exp\left({-\displaystyle
\frac{\cal A}{k_BT}}\right)$ appearing in Eq.~\ceq{psifixlen}, but may
still give a relevant contribution to the whole path integral if their
number is overwhelming with respect to the conformations of minimal
energy.

Let us now discuss the
boundary conditions which appear in the path integrals of
Eq.~\ceq{psifixlen}.
Since the interactions are not relevant in the
present context, we will consider just ideal chains.
It is also convenient to reformulate
Eq.~\ceq{psifixlen}
in terms
of the real variables $x(t,s)$ and $y(t,s)$:
\beq{
\Psi^{2d}
=\int_{\mathbf R(t_f,s)=
\mathbf R_f(s)\atop
\mathbf R(t_i,s)=
\mathbf R_i(s)
}
{\cal D}x(t,s){\cal
  D}y(t,s) e^{-\frac{{\cal A}_0}{k_BT}}\delta\left(
(\partial_sx)^2+(\partial_sy)^2-1
\right)
}{distxy}
where 
\beq{
{\cal A}_0=\frac{M}{2L} \int_{t_i}^{t_f}\int_0^Lds\dot \mathbf R^2(t,s)
}{realaction}
Apart from the boundary
conditions:
\begin{eqnarray}
\mathbf R(t_f,s)&=&\mathbf R_f(s)\label{sss}\\
\mathbf R(t_i,s)&=&\mathbf R_i(s)\label{ttt}
\end{eqnarray}
and
\beq{\mathbf R(t,0)=(l_1\cos\varphi_1,l_1\sin\varphi_1)}{fdfdf}
one could also add the requirement that the chain forms a closed
loop. This further condition is implemented by the constraint:
\beq{
\mathbf R(t,0)=\mathbf R(t,L)
}{closedloopcond}
Eq.~\ceq{closedloopcond} constrains the two ends of the chain,
occurring when $s=0$ and $s=L$, to coincide with the fixed point
$(l_1\cos\varphi_1,l_1\sin\varphi_1)$ at any instant $t$.
It is cumbersome to keep track of all these boundary conditions
within the path integral,  so that it becomes preferable to
expand the fields $\mathbf R(t,s)$ around a classical background:
\beq{
\mathbf R(t,s)=\mathbf R_{cl}(t,s)+\mathbf R_q(t,s)
}{decomp}
We first concentrate ourselves on the background.
Due to the fact that there we assumed that
the chain is ideal, 
the classical fields $\mathbf R_{cl}(t,s)$ 
can be chosen among those which obey the free equations of
motion associated to the action ${\cal A}_0$ of Eq.~\ceq{realaction}:
\beq{
\ddot\mathbf R_{cl}(t,s)=0
}{clexteqmot}
The above equation should be solved in such a way that its solutions satisfy
the boundary conditions  
(\ref{sss})~--~(\ref{fdfdf}), but not the constraint \ceq{constr},
because this is  taken into account separately
by the $\delta$ functions appearing
Eq.~\ceq{distxy}. 
Explicitly, one may write the general expression of $\mathbf R_{cl}(t,s)$
as
follows:
\beq{
\mathbf R_{cl}(t,s)=\frac{t_f\mathbf R_i(s)
-t_i\mathbf R_f(s)}{t_f-t_i}+\frac t{t_f-t_i}(\mathbf
R_f(s)-\mathbf R_i(s))
}{clsolwitcon}
The boundary condition (\ref{fdfdf}) is fulfilled by requiring that
$\mathbf R_i(0)=\mathbf R_f(0)=(l_1\cos\varphi_1,l_1\sin\varphi_1)$.
If one wishes to add closed loop condition \ceq{closedloopcond}, one
has to ask additionally that:
\beq{
\mathbf R_\tau(0)=\mathbf R_\tau(L)=\mathbf R_0(0)=\mathbf R_0(L)=
(l_1\cos\varphi_1,l_1\sin\varphi_1)
}{condsonclfie}
Let's now discuss the
component $\mathbf R_q(t,s)$ appearing in Eq.~\ceq{decomp}. It describes
the statistical fluctuations around the classical background
$\mathbf R_{cl}(t,s)$. Since the non-trivial boundary conditions are
already taken into account by the background,
$\mathbf R_{q}(t,s)$ has a trivial behavior at the boundary:
\beq{
\mathbf R_q(t_i,s)=\mathbf R_q(t_f,s)=0
}{boundcondrq}

After having split the fields $\mathbf R(t,s)$ as in
Eq.~\ceq{decomp}, the probability distribution \ceq{distxy} becomes:
\begin{eqnarray}
\Psi^{2d}
&=&\exp\left[-
\frac{{\cal A}_0(\mathbf R_{cl})}{k_BT}
\right]\nonumber\\
&\!\!\!\!\!\!\!\!\!\!\!\!\!\!\!\!\!\!\!\!\!\!\!\!\!\!\!\!\!\!\!\!\!\!\!\!\!
\times&\!\!\!\!\!\!\!\!\!\!\!\!\!\!\!\!\!\!\!\!\!\!
\int{\cal D}\mathbf R_q(t,s)\exp\left[
-\frac{{\cal A}_0(\mathbf R_{q})}{k_BT}
\right]\delta\left((\partial_s\mathbf R_{cl}^2)+
(\partial_s\mathbf R_q^2)+2\partial_s\mathbf R_{cl}\cdot 
\partial_s
\mathbf
R_q-1
\right)
\label{acthclaq}
\end{eqnarray}
where ${\cal A}_0(\mathbf R_{cl})$ contains just the classical field
conformations:
\beq{
{\cal A}_0(\mathbf R_{cl})=\frac M{L(t_f-t_i)}\int_0^Lds\left(
\mathbf R_f(s)-\mathbf R_i(s)
\right)^2
}{clafieham}
while the statistical
fluctuations are in ${\cal A}_0(\mathbf R_{q})$:
\beq{
{\cal A}_0(\mathbf R_q)=\frac M{2L}\int_{t_i}^{t_f}
dt\int_0^Lds\dot\mathbf R_q^2 
}{hamhq}
After introducing a Lagrange multiplier $\lambda(t,s)$, it is possible
to write in Eq.~\ceq{acthclaq} as follows:
\begin{eqnarray}
\Psi^{2d}
&=&\exp\left[-
\frac{{\cal A}_0(\mathbf R_{cl})}{k_BT}
\right]\nonumber\\
&\!\!\!\!\!\!
\times&\!\!\!\!\!\!
\int{\cal D}\mathbf R_q(t,s)\exp\left[
-\frac{{\cal A}_0(\mathbf R_{q})}{k_BT}
\right]\nonumber\\
&\!\!\!\!\!\!\!\!\!\!\!\!\!\!\!\!\!\!
\times&\!\!\!\!\!\!\!\!\!\!\!
\exp\int_{t_i}^{t_f} dt\int_0^L ds\left[-
\lambda(t,s)\left((\partial_s\mathbf R_{cl})^2+
(\partial_s\mathbf R_q)^2+2\partial_s\mathbf R_{cl}\cdot \partial_s
\mathbf R_q-1
\right)\right]
\label{acthclaqlagmult}
\end{eqnarray}
After the field splitting of
Eq.~\ceq{decomp}, one has to deal only with the statistical fluctuations
$\mathbf R_q$, whose boundary conditions are trivial. In this way it
becomes in principle possible to eliminate the
$\mathbf R_q$'s from
the path integral \ceq{acthclaqlagmult}
by performing a gaussian
integration. Finally, let us discuss when it is convenient to split
the fields as in Eq.~\ceq{decomp}.
We know already from Section~\ref{sec:class} that 
the constraint \ceq{constr} forces the identity
$\mathbf
R_f(s)-\mathbf R_i(s)=0$ in Eq.~\ceq{clsolwitcon} as
in the case of the classical solutions \ceq{xclone}
and \ceq{xcltwo}.
For this reason, if the final and initial conformations $\mathbf
R_f(s)$ and $\mathbf R_i(s)$ are chosen to be very different,
it is licit to expect
that the statistical fluctuations $\mathbf R_q(t,s)$ will not be small,
since they are needed to restore the constraint \ceq{constr}. Thus,
from a perturbative point of view,
it is possible to consider the fluctuations  $\mathbf R_q(t,s)$ as
small perturbations only if $\mathbf R_f(s)\sim\mathbf R_i(s)$, i.~e.
\beq{
|\mathbf R_f(s)-\mathbf R_i(s) |<\epsilon
}{pertcond}
with $\epsilon$ being a small constant parameter.
\section{The three dimensional case}
In this Section we consider a chain of $N$ segments $P_iP_{i-1}$ of
fixed lengths $l_i$,
$i=2,\ldots,N$, in three dimensions. Using spherical coordinates, the
positions of the end points of the segments
$P_i(t)=(x_i(t),y_i(t),z_i(t))$ are 
given by:
\beq{
\begin{array}{rcl}
x_n(t)&=&\sum_{i=1}^Nl_i\cos\varphi_i(t)\sin\theta_i(t)\\
y_n(t)&=&\sum_{i=1}^Nl_i\sin\varphi_i(t)\sin\theta_i(t)\\
z_n(t)&=&\sum_{i=1}^Nl_i\cos\theta_i(t)
\end{array}\qquad\qquad n=1,\ldots,N
}{posns3d}
For simplicity,
the kinetic energy $K_{disc}^{3d}$ of the system will be computed 
in the particular case in which the chain is attached at the origin
of the coordinates, i. e. 
\beq{
P_1=(0,0,0)\qquad\qquad l_1=\dot l_1=0
}{condirufdi}
After
a long but 
straightforward calculation. the result is:
\begin{eqnarray}
K_{disc}^{3d}&=&\sum_{n=1}^N\sum_{k=1}^{n-1}\frac
{m_n}2l^2_{n-k+1}\dot\varphi_{n-k+1}\sin^2\theta_{n-k+1}
+\sum_{n=1}^{N}\sum_{k=1}^{n-1}\frac{m_n}2l^2_{n-k+1}\dot\theta^2_{n-k+1}
\nonumber\\
&+&\sum_{n=1}^{N}\sum_{k=1}^{n-1}\sum_{i=2}^{n-k}m_nl_il_{n-k+1}\left[
\dot\varphi_i\dot\varphi_{n-k+1}\sin\theta_i\sin\theta_{n-k+1}\cos\left(
\varphi_{n-k+1}-\varphi_i
\right)
\right.\nonumber\\
&+&
\dot\theta_i\dot\varphi_{n-k+1}\cos\theta_i\sin\theta_{n-k+1}
\sin\left(
\varphi_i-\varphi_{n-k+1}
\right)\nonumber\\
&+&\dot\varphi_i\dot\theta_{n-k+1}\sin\theta_i\cos\theta_{n-k+1}
\sin\left(
\varphi_{n-k+1}-\varphi_i
\right)\nonumber\\
&+&\left.\dot\theta_i\dot\theta_{n-k+1}\left(
\cos\theta_i\cos\theta_{n-k+1}\cos\left(
\varphi_{n-k+1}-\varphi_i
\right)+\sin\theta_i\sin\theta_{n-k+1}
\right)\right]\label{kdisc3d}
\end{eqnarray}
It is now possible to pass to the limit of a continuous chain
$a\longrightarrow 0$ and $N\longrightarrow +\infty$. Making
the same assumptions of uniform length and mass distributions as in
Eqs.~\ceq{unifass} and \ceq{liallequal}, we get after some algebra:
\begin{eqnarray}
\!\!\!\!\!\!\!\!\!\!K^{3d}&=&\frac ML\int_0^Lds(L-s)\int_0^s dv\left[
\dot\varphi(t,v)\dot\varphi(t,s)\sin\theta(t,v)\sin\theta(t,s)
\cos\left( \varphi(t,s)-\varphi(t,v)\right) \right. \nonumber\\
&+&
\dot\theta(t,v)\dot\varphi(t,s)\cos\theta(t,v)\sin\theta(t,s)\sin\left
( \varphi(t,v)-\varphi(t,s)\right)\nonumber\\
&+&\dot\theta(t,s)\dot\varphi(t,v)\sin\theta(t,v)\cos\theta(t,s)\sin\left(
\varphi(t,s)-\varphi(t,v)
\right)\nonumber\\
&+&\left.\dot\theta(t,v)\dot\theta(t,s)\left(
\cos\theta(t,v)\cos\theta(t,s)\cos\left(
\varphi(t,s)-\varphi(t,v)
\right)
+\sin\theta(t,v)\sin\theta(t,s)
\right)
\right]\label{kinene3dformone}
\end{eqnarray}
This is the three dimensional analog of Eq.~\ceq{hamfrecha}.

The kinetic energy of  \ceq{kinene3dformone} is complicated, but
from the lesson of the two dimensional case we know how to simplify it. 
First of all, we note that the kinetic energy of the discrete chain may
be written in terms of the cartesian coordinates
\ceq{posns3d} as follows:
\beq{
K_{disc}^{3d}=\sum_{n=2}\frac {m_n}2(\dot x_n^2+
\dot y_n^2+\dot z_n^2
)
}{kdisc3dcartcoord}
where $x_n$, $y_n$ and $z_n$ have been defined in Eq.~\ceq{posns3d}.
The sum over $n$ starts from $2$ because one end of the chain
coincides with the origin of
the axes, so that $l_1=0$.
Of course, due to the condition that each segment has a fixed length
$l_i$, Eq.~\ceq{kdisc3dcartcoord} must be completed by the following
constraints:
\beq{
(x_n-x_{n-1})^2+(y_n-y_{n-1})^2+(z_n-z_{n-1})^2=l_i^2\qquad\qquad
  n=2,\ldots,N 
}{constrthreedone}
At this point we have two choices. Either we keep the
kinetic energy in the simple form of Eq.~\ceq{kdisc3dcartcoord}
at the price of having to deal with the constraints
\ceq{constrthreedone}, or we solve those constraints using spherical
coordinates $l_i,\theta_i,\phi_i$,
as it has been done in Eq.~\ceq{posns3d}. In the latter case, we have the
complicated expression of the kinetic
energy of Eq.~\ceq{kinene3dformone} and there is the problem of
defining a path integration over spherical coordinates.
In the continuous limit, the situation does not change substantially.
After performing the continuous limit following the prescriptions
of Section~\ref{sec2},
the
kinetic energy of Eq.~\ceq{kdisc3dcartcoord} and the
 constraints \ceq{constrthreedone} are respectively replaced by:
\beq{
K^{3d}=\frac{M}{2L}\int_0^Lds\left[
(\partial_t x(t,s))^2+(\partial_t y(t,s))^2+(\partial_t z(t,s))^2
\right]
}{kcontthreed}
and
\beq{
(\partial_sx(t,s))^2+(\partial_sy(t,s))^2+(\partial_sz(t,s))^2=1
}{constrthreecontcase}
The constraint \ceq{constrthreecontcase} can be eliminated
by introducing spherical coordinates $\theta(t,s),\varphi(t,s)$:
\begin{eqnarray}
x(t,s)&=&\int_0^s
du\cos\varphi(t,u)\sin\theta(t,u)\label{contcoordtransfone}
\\
y(t,s)&=&\int_0^s du\sin\varphi(t,u)\sin\theta(t,u)
\label{contcoordtransftwo}
\\
z(t,s)&=&\int_0^s du\cos\theta(t,u)
\label{contcoordtransfthree}
\end{eqnarray}
If one makes the coordinate substitutions of
Eqs.~(\ref{contcoordtransfone}--\ref{contcoordtransfthree}) in the
kinetic energy \ceq{kcontthreed} and applies
the formula \ceq{usefor}, one arrives exactly at the expression of the
kinetic energy 
\ceq{kinene3dformone}. 
Thus, Eq.~\ceq{kinene3dformone} and Eq.~\ceq{kcontthreed} together
with the constraint \ceq{constrthreecontcase} are equivalent.

To construct of the probability distribution $\Psi^{3d}
$ in three dimensions 
we choose the approach in which the 
the coordinates
  $x(t,s),y(t,s)$ and $z(t,s)$ are independent and the
  right number of degrees of freedom is restored by the condition
  \ceq{constrthreecontcase}. 
%
The result is similar to that of the two dimensional case:
\beq{
\Psi^{3d}
=
\int_{\mathbf R(t_f,s)=\mathbf R_f(s)\atop
\mathbf R(t_i,s)=\mathbf R_i(s)
}{\cal D}\mathbf R(t,s)\exp\left\{-\displaystyle\frac{{\cal A}_0^{3d}}{k_BT}
\right \}
\delta((\partial_s\mathbf R)^2-1)
}{finalthreedparfun}
where
\beq{{\cal A}_0^{3d}
=\frac{M}{2L}\int_{t_i}^{t_f}dt\int_0^Lds\left[
(\partial_t x(t,s))^2+(\partial_t y(t,s))^2+(\partial_t z(t,s))^2
\right]=\frac M{2L}\int_{t_i}^{t_f}dt\int_0^Lds \dot \mathbf R^2
}
{fdsfgf}
The above expression of the probability distribution has been obtained
assuming that one end of the chain is attached at the origin according
to the condition of Eq.~\ceq{condirufdi}. However, 
using translational invariance, it is easy to check that
Eqs.~\ceq{finalthreedparfun} and \ceq{fdsfgf} remain valid also if the
chain has one of its ends fixed at any other given point $\mathbf
R_{fixed}$,
 so that to
complete the set of boundary conditions of Eq.~\ceq{finalthreedparfun}
we may add the following requirement:
\beq{
\mathbf R(t,0)=\mathbf R_i(0)=\mathbf R_{fixed}(0)
}{additgreq}
$\mathbf R_{fixed}(0)$ being a fixed point.
%

\section{Chains with constant angle of bending}
\begin{figure}
\centering
\includegraphics[width=8cm]{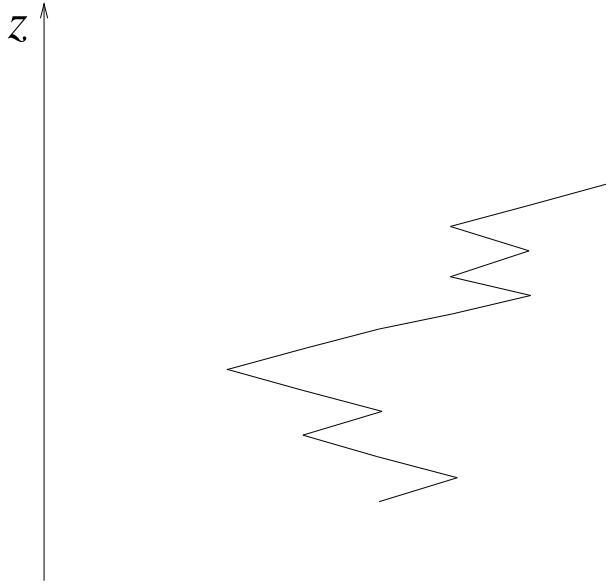}
\caption{Example of motion of a chain whose segments are constrained
  to form a fixed angle $\alpha$ with the $z-$axis. In the figure
  $\alpha=30^\circ$ } \label{motconang} 
\end{figure}

The approach presented above in order to treat the dynamics of random
chains has some interesting variants which we would like to discuss in
this Section.
To this purpose, we choose the formulation in which the positions of
the ends of the segments composing the chain are given in cartesian
coordinates. As we have already seen, in this way the expression of
the kinetic energy $K_{disc}^{3d}$ is simply:
\beq{K_{disc}^{3d}=\sum_{n=1}^N\frac{m_n}2 (\dot
  x_n^2+\dot y_n^2+\dot z_n^2)}{kthreedrewrite}
However, one has to take into account also  the constraints:
\beq{
(x_n-x_{n-1})^2+(y_n-y_{n-1})^2+(z_n-z_{n-1})^2=l_n^2\qquad n=2,\ldots,N
}{constrewrite}
We assume as before that all segments have the same fixed length
$l_n=a$, but additionally we require that:
\beq{
(z_n-z_{n-1})^2=b^2\le a^2
}{addrequir}
This implies that the projection of each segment onto the $z-$axis has
length $\pm b$, so that the segments are bound to form with the
$z-$axis the fixed angles $\alpha_1=\alpha$ or $\alpha_2=
(\pi-\alpha)$ defined by 
the relations:
\beq{\cos\alpha_1=+\frac ba\qquad\qquad \cos\alpha_2=-\frac
  ba}{anglealphadef} 
Clearly, in both cases the constraints \ceq{constrewrite} and
\ceq{addrequir} may be rewritten as follows:
\beq{
\frac{(x_n-x_{n-1})^2}{b^2}+\frac{(y_n-y_{n-1})^2}{b^2}=\frac
1{\cos^2\alpha}-1\qquad\qquad n=2,\ldots,N 
}{constralpha}
We suppose now that 
 only the angle
$\alpha_1$ is allowed,
so that the
chain cannot make turns in the $z$ direction.
An example of a conformation of a chain
satisfying these assumptions is given in Fig~\ref{motconang}.
The constraints
\ceq{constralpha} are solved by choosing spherical coordinates, in
which however the angles $\theta_n$ formed by the segments with the
$z-$axis are always equal to $\alpha$:
\begin{eqnarray}
x_n(t)&=&\sum_{i=1}^nl_i\cos\varphi_i(t)\sin\alpha\label{varonedf}\\
y_n(t)&=&\sum_{i=1}^nl_i\sin\varphi_i(t)\sin\alpha\label{vartwodf}\\
z_n(t)&=&\sum_{i=1}^nl_i\cos\alpha=n\cos\alpha\label{coordsab}
\end{eqnarray}
As we see from the above equation, each segment is left  only
with the freedom of 
rotations around the $z-$direction, corresponding to the angles
$\varphi_i(t)$. Moreover, the total length of the chain is always
$L=Na$, but now also the total height $h$ of the trajectory along the
$z-$axis is fixed:
\beq{h=Nb}{totalheightz}
At this point, we pass to the continuous limit, this time taking as
parameter describing the trajectory of the chain the variable $z$
instead of the arc-length $s$. Due to the
last of Eqs.~\ceq{coordsab}, the $z-$components of the velocities are
always zero: 
\beq{\dot z_n(t)=0}{constzdot}
As a consequence, we are left with a two dimensional
problem, which may be treated in exactly the same way as we treated the two
dimensional chain of Section~\ref{sec2}. 
The only difference is that Eqs.~\ceq{carpol2} should be replaced by
Eqs.~\ceq{varonedf} and \ceq{vartwodf} and the constraints have a
slightly different form.
Following the same procedure presented in Section~\ref{sec2} we find
after a few calculation the expression of 
the kinetic energy:
\begin{eqnarray}
K^{3d}_{\alpha}&=&\tan^2\alpha\int_0^hdz\int_0^zdz_1\int_0^{z_1}dz_2
\rho_m(t,z)\rho_l(t,z-z_1)\rho_l(z_2)\nonumber\\
&\times&
\dot\varphi(t,z-z_1)\dot\varphi(t,z_2)
\cos( \varphi(t,z-z_1) -\varphi(t,z_2))\label{kcont3dalphaone}
\end{eqnarray}
and of the constraint 
\ceq{constralpha}:
\beq{
(\partial_z x)^2+(\partial_z y)^2=\tan^2\alpha
}{constralphacont}
Apart from the appearance of the factor $\tan^2\alpha$ and the
choice of the height $z$ instead of the arc-length $s$,
Eqs.~\ceq{kcont3dalphaone} and \ceq{constralphacont}
are identical to Eqs.~\ceq{Tcont} and \ceq{constrthreedone} in
the limit $l(t)=\dot l(t)=0$. It is now not difficult to show that the
probability distribution $\Psi^{3d}_{\alpha}
$ is given by:
\beq{
\Psi^{3d}_{\alpha}
=
\int{\cal D}x(t,z){\cal D}y(t,z) \exp\left\{-\frac{{\cal A}_{0,\alpha}}{k_BT}
\right\}\delta(
(\partial_z x)^2+(\partial_z y)^2-\tan^2\alpha
)
}{3ddirfin}
where
\beq{
{\cal A}_{0,\alpha}=\tan^2\alpha\int_{t_i}^{t_f}dt\int_0^hdz\left[
\dot x^2+\dot y^2
\right]
}{djfjskdflds}

\begin{figure}
\centering
\includegraphics[width=8cm]{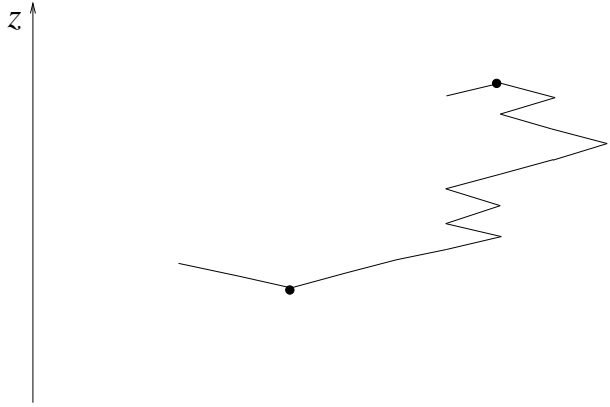}
\caption{Example of motion of a chain whose segments are constrained
  to form fixed angles $\alpha$ or
$\pi-\alpha$
 with the $z-$axis. In the figure
  $\alpha=30^\circ$.  Turning points are emphasized by means of black points.}
  \label{turnedchain}  
\end{figure}

At this point we discuss briefly the case in which both angles
$\pi-\alpha$ and 
$\alpha$ are allowed. In this situation, the trajectory of the chain
may have 
turns. An example of motion of this kind is given in
Fig~\ref{turnedchain}.
The constraints \ceq{constrewrite} and \ceq{addrequir} remain
unchanged, but the coordinate $z$ cannot be chosen as a valid
 parameter of the trajectory of the chain and
one has to come back to the arc-length $s$. 
The most serious problem
occurs due to the fact that the variables
$z_n(t)$ are not continuous functions of the time, since
each $z_n(t)$ is allowed to jump
discretely between the two discrete values $+b$ and $-b$, corresponding
to the angles $\alpha$ and $\pi-\alpha$ respectively. It is
therefore difficult to define the components $\dot z_n$ of the
velocities of the ends of the segments and thus their contribution to
the  kinetic energy. 
Let us note that this problem affects only the $z$ degrees of freedom.
The degrees of freedom $x_n(t)$
and $y_n(t)$
of the chain remain continuous functions of $t$ despite the jumps
of the $z_n$'s. This fact can be easily verified looking at the
definition of $x_n(t)$
and $y_n(t)$
in Eqs.~\ceq{varonedf} and \ceq{vartwodf}. Since 
$\sin(\pi-\alpha)=\sin\alpha$, both the  $x_n(t)$'s
and $y_n(t)$'s are not affected by the jumps of the angle
$\alpha\longleftrightarrow \pi-\alpha$. The situation 
simplifies only if the 
chain has no interactions in which the $z$
variable is 
involved. The reason is that in the kinetic energy and in the constraints
given by Eqs.~(\ref{kthreedrewrite}--\ref{addrequir})
the degrees of
freedom connected to the motion along the $z-$directions
are decoupled from the other degrees of freedom and may be ignored. 
As a consequence, in
absence of $z-$dependent interactions, the  
difficulties related to the motion along the $z-$direction disappear
and once again the problem reduces to that 
the two dimensional chain treated in
Section~\ref{sec2}. Since the constraints are always those of
Eqs.~(\ref{kthreedrewrite}--\ref{addrequir}) one may proceed
as in the case of fixed angle $\alpha$. As a result, one finds that
the final probability distribution is of the form:
\begin{eqnarray}
\Psi^{3d}_{\alpha,\pi-\alpha}
&=&\nonumber\\
&&\!\!\!\!\!\!\!\!\!\!\!\!\!\!\!\!\!\!\!\!\!\!\!\!\!\!\!\!
\!\!\!\!\!\!\!\!\!\!\!\!\!\!\!\!\!\!\!\!\!\!\!\!\!\!\!\!\!\!\!\!
{\cal C}\int{\cal D}x(t,s){\cal D}y(t,s) \exp\left\{-\frac{{\cal
    A}_{0,\alpha,\pi-\alpha}}{k_BT} 
\right\}\delta(
(\partial_s x)^2+(\partial_s y)^2-\tan^2\alpha
)
\label{3ddirfiwn}
\end{eqnarray}
where
\beq{
{\cal
  A}_{0,\alpha,\pi-\alpha}=\sin^2\alpha\int_{t_i}^{t_f}
dt\int_0^Lds\left[  
\dot x^2+\dot y^2
\right]
}{djfjskdfldws}
and ${\cal C}$ is a constant containing the result of the integration
over the decoupled $z$ degrees of freedom. Let us note in
Eqs.~\ceq{3ddirfiwn} and \ceq{djfjskdfldws} the appearance of the
factor $\sin^2\alpha$ in the action instead of $\tan^2\alpha$ and the
replacement of $z$ with the arc-length $s$ as the parameter of the
trajectory of the chain.
\section{Conclusions}
The dynamics of a random chain has been discussed both from the
classical and statistical point of view. The classical energy of the
chain has been derived in two and three dimensions. We have mainly
concentrated ourselves on the computation of the kinetic energy
because the addition of the contribution of the interactions to the
total energy is straightforward, see Eqs.~\ceq{eneext} and
\ceq{eneint}.
The kinetic energy on the contrary has a complicated and
non-Markoffian expression as shown by Eq.~\ceq{tdiscr} in
two-dimensions and by Eq.~\ceq{kdisc3d} in three dimensions.
After passing to the continuous limit discussed in Section II, one
term disappears from the kinetic energy. This is the reason for which
in the continuous case the classical equations of
motion have simple solutions of the form given in
Eqs.~\ceq{xclone}--\ceq{xcltwo}. Let us note that the passage to the
continuous limit is straightforward and does not need mathematical
subtleties as the analogous limit by which the Edwards model is
obtained in the statistical mechanics of random chains.

The stringy approach to the dynamics of random chains is formulated in
Section IV. 
One would naively expect that the final model is related with
non-relativistic string theories, such as those derived in
Refs.~\cite{nonrelst}, but this is not the case.
In two and three dimensions the probability distributions
$\Psi^{2d}$ and $\Psi^{3d}$ are respectively given in
Eqs.\ceq{psifixlen} and \ceq{finalthreedparfun}--\ceq{fdsfgf}.
As it is possible to see from these equations, the path integral sums
which provide the expressions of $\Psi^{2d}$ and
$\Psi^{3d}$ have the form of a $O(n)$ nonlinear sigma model on a two
dimensional world-sheet, where $n=2,3$ depending on the dimensionality
of the physical space in which the chains fluctuate.
The difference with respect to the nonlinear sigma model is that here
the derivatives of the fields with respect to the arc-length $s$
are subjected to
the condition $(\partial_s\mathbf R)^2=1$. In standard nonlinear sigma models
it is instead  the modulus of the
fields themselves to be constrained. The presence of a complicated
constraint  and the absence of a
small parameter 
which could be used to start a perturbative expansion
complicate the
computation of the 
probability functions. For this reason, in Section IV it has been
proposed the field splitting \ceq{decomp}. This splitting provides a
convenient way to deal with the boundary conditions satisfied by the
fields $\mathbf R(t,s)$ and also allows a perturbative treatment
provided the initial and final conformations $\mathbf R_i(s)$ and
$\mathbf R_f(s)$ do not differ very much in the sense of
Eq.~\ceq{pertcond}. Alternatively, 
the statistical fluctuations
$\mathbf R_q(t,s)$ may  be eliminated from the partition function of
Eq.~\ceq{acthclaqlagmult} with a Gaussian integration
\footnote{F. Ferrari
would like to thank L. Cugliandolo for suggesting this possibility.},
but then one 
ends up with a complicated two dimensional field theory in which the
Lagrange multipliers $\lambda(t,s)$ are coupled together with the
classical background $\mathbf R_{cl}(t,s)$. Another possibility in
order to simplify the model
consists in relaxing the constraint $(\partial_s\mathbf R)^2=1$,
imposing for instance the weaker condition
$\frac{1}{TL}\int_{t_i}^{t_f}dt\int_0^Lds(\partial_s\mathbf R)^2 =1$.
This constraint requires that the average length of the
chain in the interval of
time $t_i\le t\le t_f$ is $L$. In this case
the Lagrange multiplier $\lambda$ does no longer depend on $t$ and $s$ and
calculations become easier. It would also be interesting to study the
probability functions $\Psi^{2d}$ and $\Psi^{3d}$ in the limit
$t_f\longrightarrow +\infty $ and $t_i\longrightarrow -\infty$ or to
sum them over all possible values of the initial and final
conformations $\mathbf R_i(s),\mathbf R_f(s)$.

Finally, chains with fixed angles have been discussed in Section
VI. Our approach is valid only if the chain has no turning points.
If there are turning points 
 the kinetic energy is not well defined, because the variable
$z(t,s)$ is no longer a continuous function and thus its time
derivative becomes a distribution. One way for adding to our treatment
turning points as those of
Fig.~\ref{turnedchain} is to replace the variable $z$ with a
stochastic variable which is allowed to take only discrete
values. Another way is to look at turning points as points in which
the chain bounces against an invisible obstacle. A field theory describing
a one-dimensional chain with such kind of non-holonomic constraints
has been already derived
in Refs.~\cite{arodz}.
The problem of turning points is currently work in progress,
as well as the possibility of imposing topological constraints on the
trajectory of the chain.


\begin{thebibliography}{99}
\bibitem{doiedwards} M. Doi and S.F. Edwards, The Theory of Polymer Dynamics
  (Clarendon Press, Oxford, 1986).  

\bibitem{rouse} P. E. Rouse, {\it J. Chem.
    Phys.} {\bf 21} (1953), 1272.  

\bibitem{zimm} B. H. Zimm, {\it J. Chem.
    Phys.} {\bf 24} (1956), 269.  

\bibitem{yamakawa} H. Yamakawa, {\it Ann. Rev. Phys. Chem.} {\bf 35}, 23, 1984

\bibitem{stockmayer} P.H. Verdier and W.H. Stockmayer, {\it J. Chem. Phys}, {\bf
    36}, 227, 1962

\bibitem{arti} A. Dua and T.A. Vilgis, {\it Phys. Rev E}, {\bf 71},  021801,
  2005 
  
\bibitem{zinn} J. Zinn - Justin, Quantum Field Theory and Critical Phenomena,
  Clarendon Press, Oxford, 2002,

\bibitem{kleinertpi} H. Kleinert, {\em Path
    Integrals in Quantum Mechanics, Statistics, Polymer Physics, and Financial
    Markets}, (World Scientific Publishing, 3nd Ed., Singapore, 2003).

\bibitem{nlsigma} M. Gell-Mann and M. L\'evy, {\it Nuovo Cim.} {\bf 16}
  (1965), 705; B. W. Lee, {\it Chiral Dynamics}, Gordon and Breach, 1972.

\bibitem{edwa} S. Edwards, {\it Proc. Phys. Soc.} {\bf 91} (1967), 513; {\it
    J. Phys. A}1 (1968), 15.  

\bibitem{tomapier} W. Tomaszewski and P.
  Pieranski, {\it New Jour. Phys.} {\bf 7} (2005), 45.  H. Kleinert, {\it
    Gauge Fields in Condensed Matter}, Vol 1, (World Scientific, 1990).
\bibitem{kleinertgf} H. Kleinert, {\it Gauge Fields in Condensed Matter}, Vol
  1, (World Scientific, 1990).  \bibitem{nonrelst} J. Gomis and H. Ooguri,
  {\it J. Math. Phys.} {\bf 42} (2001), 3127 [arXiv: hep-th/0009181]; U. H.
  Danielsson, A. Guijosa and M. Kruczenski, {\it JHEP} {\bf 0010} (2000), 020
  [arXiv: hep-th/0009181].  \bibitem{arodz} H. Arod\'z, P. Klimas and T.
  Tyranowski, {\it Acta Phys. Pol.} {\bf B 36} (2005). 3861; H. Arod\'z, {\it
    Acta Phys. Pol.} {\bf B 33} (2002). 1241; H. Arod\'z, {\it Acta Phys.
    Pol.} {\bf B 35} (2004). 625.
\end{thebibliography}
\end{document}